\documentclass[iop,portrait,a4paper]{emulateapj}
\usepackage{natbib}
\usepackage{xspace}
\usepackage[usenames]{color}

\newcommand{\ie}{i.e.\xspace}

\newcommand{\rwd}{r_\mathrm{WD}}
\newcommand{\dbfifteen}{\Delta m_{15}(B)}
\newcommand{\dbthirtyfive}{\Delta m_{35}(B)}

\newcommand{\sn}{SN\xspace}
\newcommand{\sne}{SNe\xspace}
\newcommand{\snia}{SN~Ia\xspace}
\newcommand{\sneia}{SNe~Ia\xspace}

\newcommand{\poslen}{r_\mathrm{0}}
\newcommand{\dislen}{r_\mathrm{D}}
\newcommand{\posvec}{\bar{r}_\mathrm{0}}
\newcommand{\disvec}{\bar{r}_\mathrm{D}}
\newcommand{\nexvec}{\bar{r}_\mathrm{next}}

\newcommand{\edit}[1]{#1}

\def\lsim{\raise0.3ex\hbox{$<$}\kern-0.75em{\lower0.65ex\hbox{$\sim$}}}
\def\gsim{\raise0.3ex\hbox{$>$}\kern-0.75em{\lower0.65ex\hbox{$\sim$}}}

\begin{document}


\title{Perturbations of \sneia lightcurves, colors and
spectral features by circumstellar dust}


\author{Rahman Amanullah and Ariel Goobar}
\affil{The Oskar Klein Center, Physics Department, Stockholm University,
    AlbaNova University Center, SE 106 192 Stockholm, Sweden}

%




\begin{abstract}
It has been suggested that multiple scattering on circumstellar dust
could explain the non-standard reddening observed in the line-of-sight
to Type Ia supernovae. In this work we use Monte Carlo simulations to
examine how the scattered light would affect the shape of optical
lightcurves and spectral features. We find that the effects on the
lightcurve widths, apparent time evolution of color excess and
blending of spectral features originating at different photospheric
velocities should allow for tests of the circumstellar dust hypothesis
on a case by case basis. Our simulations also show that for
circumstellar shells with radii $r=10^{16}-10^{19}$ cm, the lightcurve
modifications are well described by the empirical $\Delta m_{15}$
parameter and intrinsic color variations of order $\sigma_{BV}=0.05-0.1$
arise naturally. For large shell radii an excess lightcurve 
tail is expected in $B$~band, as observed in e.g. SN~2006X.
\end{abstract}





\section{Introduction}
A dramatic breakthrough in cosmology took place more than a decade ago
when the accelerated expansion of the universe was discovered through
Type Ia supernovae (\sneia) used as standardizable candles
\citep{1998AJ....116.1009R,1999ApJ...517..565P}. \sneia remain among
the best tools to investigate the energy content of the universe, as
demonstrated by recent very successful surveys of high-$z$ \sne
\citep{2006A&A...447...31A,2007ApJ...666..694W,2007ApJ...659...98R,2008ApJ...686..749K,2009ApJS..185...32K,2010ApJ...716..712A}.
However, the progress for precision cosmology is hampered by
systematic uncertainties, notably a potential drift in the \snia
luminosity and intrinsic colors with cosmic time and the impact of
changing dust extinction properties along the line of sight
\citep{2008JCAP...02..008N}. The standardization of \sneia generally
uses two parameters: the measured color, typically rest-frame $B-V$
and often one more optical color, and an empirically defined
lightcurve shape parameter, such as stretch, $s$
\citep{1997AAS...191.8504P,2005A&A...443..781G,2008ApJ...681..482C},
$x_1$ \citep{2007A&A...466...11G}, $\dbfifteen$
\citep{1993ApJ...413L.105P,1996AJ....112.2438H} or $\Delta$
\citep{1996ApJ...473...88R,2007ApJ...659..122J}. In this work we
examine how multiple scattering of supernova photons on the
surrounding material could affect both the lightcurve shapes and
colors.

While the reddening of \sneia was originally thought to be entirely
due to extinction by interstellar dust, there is an increasing body of
evidence showing that the measured color excesses to some \snia show a
steeper wavelength dependence than what is observed for reddened stars
in the Milky Way. Likewise, there is no generally accepted model that
explains the lightcurve shape variations observed in \sneia and its
correlations with peak supernova brightness
\citep{1993ApJ...413L.105P}.

Following up the work by \cite{2005ApJ...635L..33W},
\cite{2008ApJ...686L.103G} (hereafter G08) showed that multiple
scattering on circumstellar dust could potentially help explaining the
low values of the total-to-selective extinction parameter,
$R_V=A_V/E(B-V) < 3$, observed in the sight lines of near-by Type Ia
\sne
\citep{2006MNRAS.369.1880E,2008MNRAS.384..107E,2007AJ....133...58K,2008ApJ...675..626W,2008A&A...487...19N}.

A Monte Carlo ray tracing technique was used in G08 to investigate the
reddening effects in the presence of circumstellar dust and an
approximate reddening law for optical and near-IR wavelengths was
derived. \citet{2010AJ....139..120F} compared the parametrized model
in G08 with high quality $uBgriYJHK$ data from the {\em Carnegie
  Supernova Project} and found that it provided particularly good fits
for the two most reddened \sneia in their sample, SN~2005A and SN~2006X.
The success of these comparisons calls for further scrutiny of the
model. In this work we investigate the impact of multiple scattering
on the shapes of \sneia optical lightcurves. We explore how the extra
wavelength dependent ``random walk'' of photons in the presence of
scatterers affects the width of broad-band filter lightcurves. In
particular, reddened ``slow decliners'' may be compatible with the
scenario in G08.

Furthermore, also the strengths of spectroscopic features as a
function of time are likely to be affected, since a varying
path-length would lead to a superposition of several \sn phases.
Throughout the paper we point out how observations could put bounds on
the size of a hypothetical circumstellar dust shell.

\section{Circumstellar dust in the single degenerate scenario}
\label{sec:case}
In the currently favored Type Ia scenario, a white dwarf accretes mass
from a companion star until the Chandrasekhar limit is reached, at
which point instabilities lead to an explosion. However, the model
requires the accreting white dwarf to expel material accumulating on
its surface through fast stellar winds, $v_{w}\sim 10^2-10^3$
km~s$^{-1}$, to avoid the formation of a common envelope between the
white dwarf and the donor star which would prevent the explosion
\citep{1996ApJ...470L..97H,1999ApJ...517L..47H,1999ApJ...519..314H}.
Condensation of dust grains may take place in the ejected matter,
possibly also in collisions with the accreting matter. For dust grain
growth in stellar winds, the density as a function of distance
to the star surface, $r$, is generically expected to scale as $\rho_{dust}
\propto {\dot M_w \over v_{w} r^2}$, where, $v_w$, is the stellar wind
velocity and $\dot M_w$ is the mass-loss per unit time, typically of
order $10^{-7}~M_\odot~\mathrm{yr}^{-1}$ in \sneia models.

\section{Circumstellar dust destruction by \snia luminosity}
\label{sec:sublimation}
So far, we have assumed that the supernova explosion site is
surrounded by dust. However, the radiation from the supernova itself
will sublimate dust grains as these are heated up to
$T_{sub}\sim3000$~K. Thus, assuming that the entire UV-optical
luminosity of the supernova, is absorbed by surrounding dust grains,
we can compute the radius, $r_c$, where dust grains, if present at the
explosion time, would be depleted. From the radiation balance
condition:
\begin{equation}
r_c = 
\left( 
{ {L_{bol}(t)} \over {16 \pi \sigma T_{sub}} } { {Q_{abs}} \over {<Q_{IR}>} } 
\right) ^{1/2},
\label{eq:RC}
\end{equation}
where $Q_{abs}$ is the effective absorption efficiency and $<Q_{IR}>$
is the Planck-averaged emissivity at $T=T_{sub}$.
\citet{2000ApJ...537..796W} studied a similar situation, but for dust
surrounding gamma-ray bursts and found that for $2000 < T < 3000$~K,
$<Q_{IR}> \approx 0.1$ for a mixture of astronomical silicate and
graphite grains with typical sizes of 0.1 $\mu$m.
\citet{2004A&A...428..555S} have estimated the peak bolometric
luminosity of \snia to be $L_{bol} = 1.43 \cdot M_{Ni}\cdot 10^{43}$
ergs/s, where the $^{56}$Ni mass (in units of solar masses), $M_{Ni}$,
has been measured to be between 0.1 and 1.1 $M_\odot$
\citep{2000A&ARv..10..179L}. Inserting this peak luminosity ($E_\gamma
\sim 1-5$ eV) into Eq.~(\ref{eq:RC}), and assuming that all the
radiation is absorbed, we find $r_c \lsim 10^{16}$ cm ($\sim 0.003$
pc), which we will use as a rough estimate of the region where
evaporation of pre-existing dust would take place as a result of the
\sn heating. A similar estimate of the region that ought to be
depleted of dust was done by \citet{1986A&A...155..291P}. Thus, to
create a dust shell that would survive the radiation from the
supernova explosion, and using the stellar wind velocities and
mass-loss rates from Section \ref{sec:case}, the wind must start about
$10$--$100$~years prior to the explosion, i.e. $\gsim 10^{-6} M_\odot$
of mass must be expelled for any effect of circumstellar dust to be
noticeable. For SN~2005gj, a well-studied example of a supernova with
significant interaction with the circumstellar environment,
\citet{2006ApJ...650..510A} find evidence for mass-loss about 8 years
before the explosion. For significant opacity, $\tau\sim1$, the
required dust mass can be estimated from the absorption cross-sections
in \citet{2001ApJ...548..296W} and \citet{2003ApJ...598.1017D},
$\sigma_a/m_{dust} \sim 10^{4} $cm$^2/$g. Thus, for a thin dust shell
at $r \sim r_c \sim 10^{16}$ cm, this corresponds to $m_{dust} \sim
10^{-4} M_\odot$.

We also note that since $r_c \propto \sqrt{M_{Ni}}$, intrinsically
brighter explosions are likely to have larger $r_c$, i.e., thinner
circumstellar shells. In the following sections we will see that,
depending on the size of the outer shell, multiple scattering could
generate relations between brightness and lightcurve shapes and colors
qualitatively similar to the observed ranges. We note, however, that
interaction with dust in the CS material can only broaden the pristine
lightcurve, i.e., it is not possible to produce a faster decline than
what results from the original supernova explosion.

\section{Multiple scattering of photons around the supernova}
A natural effect of light scattering is that it adds flight time to
the photons, thereby affecting the observed lightcurve
\citep{2005MNRAS.357.1161P,2006MNRAS.369.1949P}. Next, we start by
generalizing the \citet{2008ApJ...686L.103G} model: we consider a
scenario where the exploding white dwarf ($\rwd\sim 10^{9}$~cm) is at
the center of a homogeneous dust shell with an outer radius, $r$
($r\gg\rwd$), and an inner radius $r_i$, where $\rwd \leq r_i \leq r$.
Note that this will not change any of the results in G08, where
$r_i\equiv\rwd\approx0$ was assumed, but it will affect the
distribution of photon flight times, i.e., the subject of this paper.

The technique in G08 is extended to investigate
the time domain by calculating the total traveled path length, $D$,
for each photon before it crosses a plane, $P$, that is perpendicular
to the line of sight between the observer and the source, and is
located at the the distance, $r$, from the source which is illustrated
in Fig.~\ref{fig:sphere}. The details for the calculation of the
photon travel times are outlined in Sec.~\ref{sec:sphere}. The
traveled photon path length, $D$, is given by
Eq.~\ref{eq:pathlength}.

The traveled photon path length is distributed in the interval $r\leq
D<\infty$, with $D=r$ for photons that do not interact with the dust
at all. The shape of the distribution of $D$ will depend on the
radius, $r$, the thickness of the shell, the scattering and absorption
properties of the dust, and the dust density. In this work we will
quantify the latter in terms of the observed color excess, $E(B-V)$,
relative to the pristine source color. For the scattering and
absorption properties we use models consisting of carbonaceous grains
and amorphous silicate grains from \citet{2001ApJ...548..296W} and
\citet{2003ApJ...598.1017D}.

\begin{figure*}[tb]
  \centering
  \includegraphics[width=\textwidth]{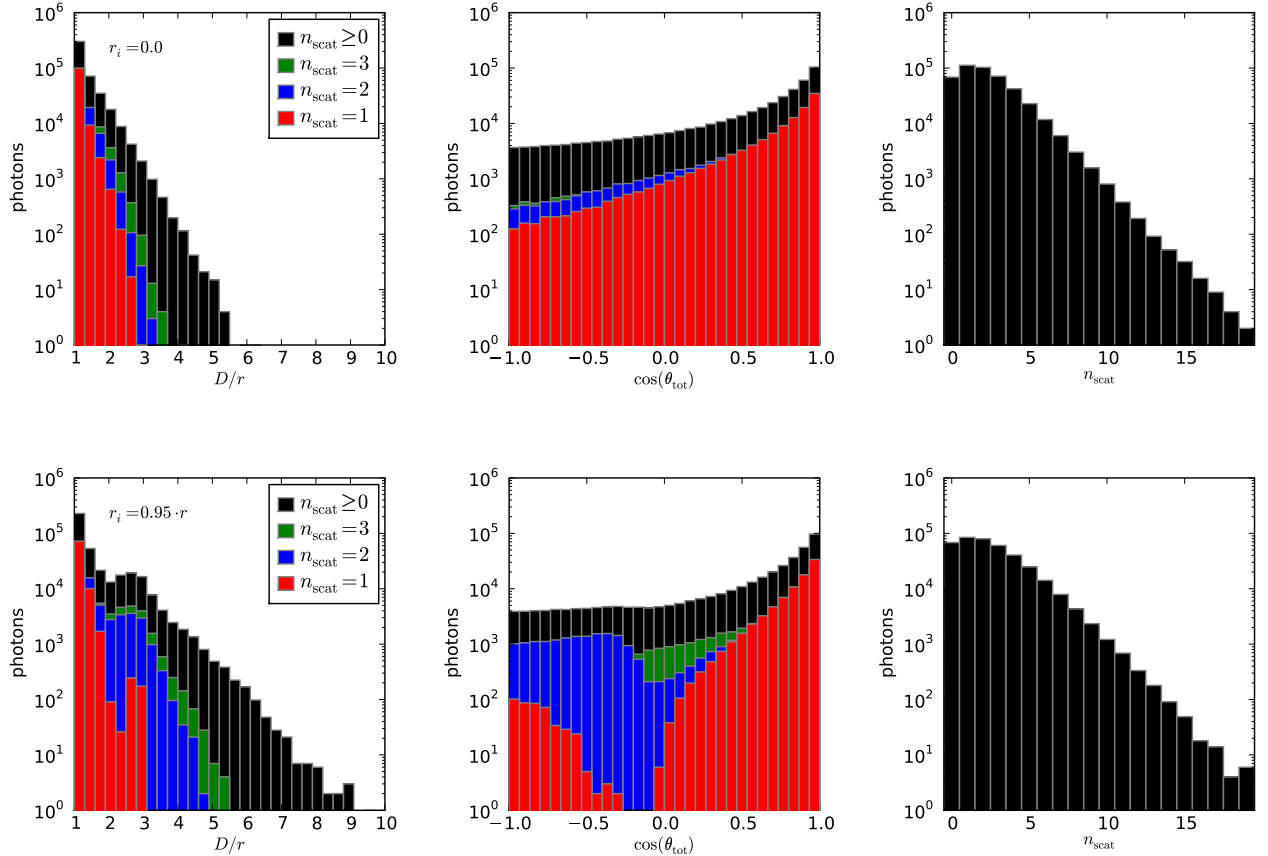}
  \caption{Monte Carlo results of two extreme circum stellar dust
    scenarios with a dust density corresponding to an observed
    $E(B-V)=0.5$. The {\em upper} and {\em lower} panels show the
    results for $r_i=0.0$ and $r_i=0.95\cdot r$ respectively. The {\em
      left} panels show the histograms of the total traveled photon
    paths $D$ (see Sec.~\ref{sec:sphere} for definition) normalized
    with the radius, $R$, of the sphere, while the {\em middle} panels
    show the histogram of the total scattering angles, and the {\em
      right} panels show the number of scatters. The local minimum
    around $D/r\sim2$ for the lower left panel is related to the lack
    of photons in the lower middle panel around
    $\cos(\theta_{\mathrm{tot}})\sim0$ with $n_\mathrm{scat}=1$.}
    \label{fig:pathdistrib}
\end{figure*}

We begin by exploring the scattering distribution for the two extreme
scenarios (a) $r_i = \rwd \approx 0$ (\ie, a homogeneous dust sphere
as in G08), and (b) $r_i = 0.95\cdot r$, \ie,
a very thin dust shell. Clearly, the scenario (a) is already
challenged by the conclusions in Section \ref{sec:sublimation},
nevertheless, it serves as a point of reference. 
The left panel of Fig.~\ref{fig:pathdistrib} shows the distribution of
$D/r$ from Monte Carlo simulations of photons with $\lambda=5000$~\AA.
Most photons travel unscattered, and for the rest, there is a rough
exponential fall-off of delay times. The dust density was
chosen to cause an observed color excess $E(B-V)\approx0.5$.
Furthermore, we are assuming that the dust grain properties correspond
to LMC-type dust, although the time distributions for MW-type dust are
very similar. 

A local minimum can be noted for case (b) around $D/r\sim2$. This can
be understood by realizing that photons from the source can only
double the path length, $D/r\sim2$, by having a total scattering angle
of $\sim\pi/2$ for this particular setup, which will take it through
more dust than for other angles. It is then likely to scatter again,
which would result in either a shorter or a longer path. The
significance of the minimum will decrease with the density of the dust
and the inner radius of the shell .


\section{Lightcurve perturbation from multiple scattering}
Since the scattering cross-section, scattering angle and albedo are
wavelength dependent, the lightcurve perturbations are expected
to vary between broad-band filters.

We use the time dependent spectral energy distribution (SED) \snia
templates from \citet{2007ApJ...663.1187H} (hereafter H07) to build up
observed broad-band lightcurves in the presence of circumstellar dust.
Thus, we make the simplifying assumption that the empirically derived
spectral template is pristine, \ie, not already affected by dust. This
is not a major limitation for this analysis, since we are here
primarily concerned with relative quantities, such as time-delays or
color excess rather than absolute times or fluxes.

Like in G08, we only consider dust grain
models from \citet{2001ApJ...548..296W} and
\citet{2003ApJ...598.1017D} for Milky-Way dust and LMC dust. In
particular, it was found in G08 that LMC dust
grains in the circumstellar medium would lead to $R_V\approx 1.65$,
while Milky-Way dust lead to $R_V\approx 2.56$. These two grain models
approximately bracket the empirically found values for $R_V$ for a
large fraction of \sneia. In both cases, the extinction law deviated
significantly from Galactic extinction \citep{1989ApJ...345..245C} in the UV
and NIR regions.

For the idealized case considered in this work, a fraction of the
photons scatter inside the dust shell and the delay time induced by
the multiple scattering will scale with $r$,
\begin{equation}
  \Delta t_r \sim {r \over c} 
  \sim 4 \cdot \left( {r \over 10^{16} {\rm cm}} \right) {\rm days}. 
  \label{eq:time-delay}
\end{equation}
Since the typical rise-time of \sneia lightcurves is about 3~weeks, and
the fall-time is somewhat longer, we expect significant effects on the
shapes of optical lightcurves for $r \gsim 10^{16}$~cm for increasing
optical depth, $\tau(\lambda)$. On the other hand, for time scales
$\Delta t_r \gg 1$ month, i.e., $r \gg 10^{17}$~cm, we expect the time
perturbations to de-correlate with the lightcurve shape parameters
currently used,
since these are designed to capture smaller time scale effects.

\section{Lightcurve shape}
Existing Type Ia supernova lightcurve fitters
\citep{1996ApJ...473...88R,2001ApJ...558..359G,2003ApJ...590..944W,%
  2005A&A...443..781G,2007A&A...466...11G,2007ApJ...659..122J,%
  2008ApJ...681..482C,2011AJ....141...19B} use different approaches to
take variations of the optical lightcurve shape into account for
standardizing \sneia. To estimate the effect on the lightcurve shape
due to circumstellar dust, we once again consider the two extreme
scenarios: (a) $r_i = \rwd \approx 0$ and (b) $r_i = 0.95\cdot r$,
introduced above.

\begin{figure*}[tb]
  \centering
  \includegraphics[width=\textwidth]{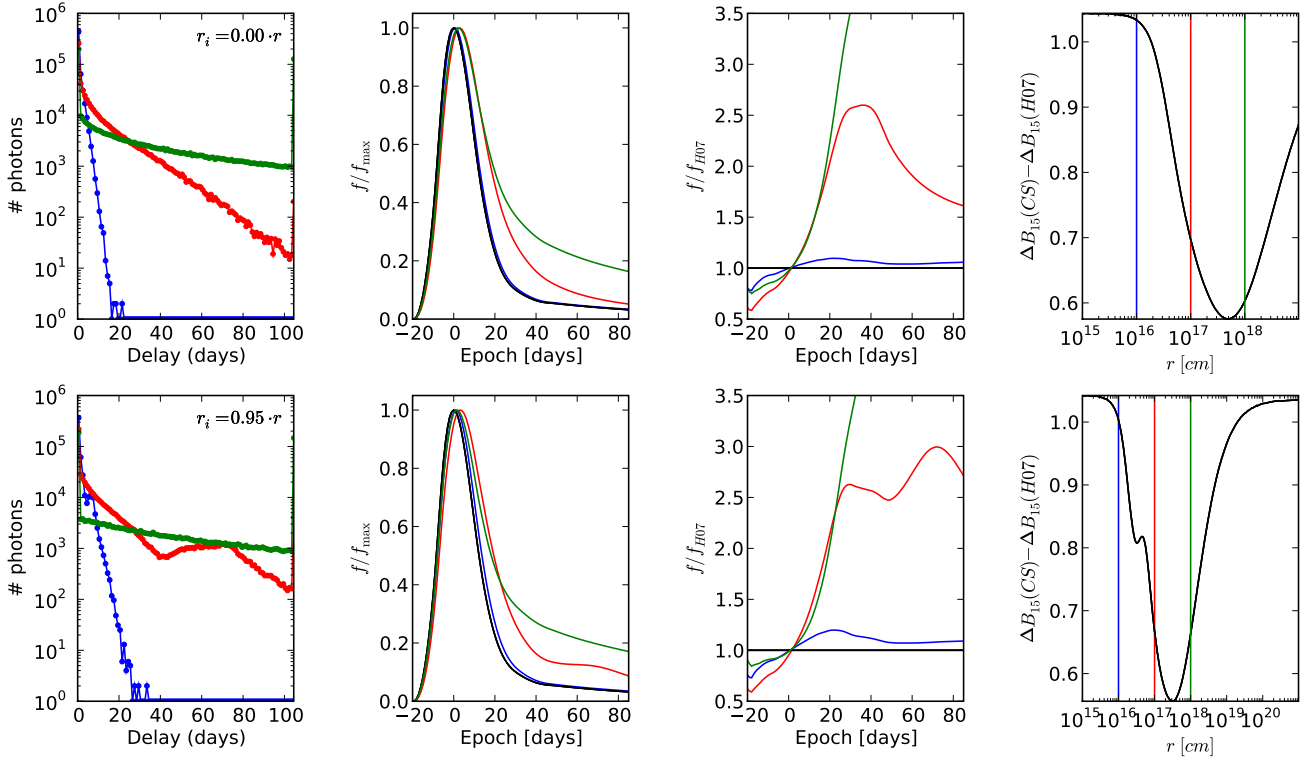}
  \caption{Simulated effects on the \snia $B$~band lightcurve from
    multiple scattering by circumstellar dust for three different
    radii: (blue,red,green) = ($10^{16}$,$10^{17}$,$10^{18}$)~cm. The
    {\em upper} row shows results for $r_i=0.0$ while the {\em lower}
    is for $r_i=0.95\cdot r$. A reddening of $E(B-V)\approx0.12$
    has been assumed for both cases.  The {\em far left} column shows
    the photon delay distribution in the $B$~band.  The {\em middle
      left} column show the corresponding $B$~band light curves after
    the source (black) SED has been convolved by the delay
    distributions. The {\em middle right} column gives the $B$
    lightcurves normalized by source lightcurve at maximum.  The {\em
      far right} column shows the $B$~band lightcurve $\dbfifteen$
    for different values of circum stellar shell radii.
    \label{fig:rdependence}}
\end{figure*}

Fig.~\ref{fig:rdependence} shows the time-delay distribution in the
$B$~band from Monte Carlo simulations as a function of the outer shell
radius $r$ for cases (a) and (b) where the LMC-type dust density was
chosen to cause an observed color excess $E(B-V)\approx0.12$.

The far left panel of this figure is closely related to the left panel
of Fig.~\ref{fig:pathdistrib}, except that the dimensionless property $D/r$
has been converted to days by assuming different physical shell radii.
First, we note that for very large outer radii, photons that scatter
on dust grains have a fairly flat delay probability with respect to
the unscattered ones in the first (zero) bin. We also note how the
effect that certain photon paths are less likely for a thin shell, as
discussed above, propagates to a sharp feature in the delay time
distribution.

Also shown in Fig.~\ref{fig:rdependence} is
 how the delayed photons perturb the pristine
$B$~band lightcurve shape. This reveals how the size of the outer radius, 
affects (primarily) the
post-max shape of the lightcurve. Notably, the largest deviation from the
pristine shape can be seen in the tail, more than a month after lightcurve
maximum, typically beyond the time range investigated for most 
reddened supernovae.

\subsection{The fall-time}
\begin{figure*}
  \centering
  \includegraphics[width=\textwidth]{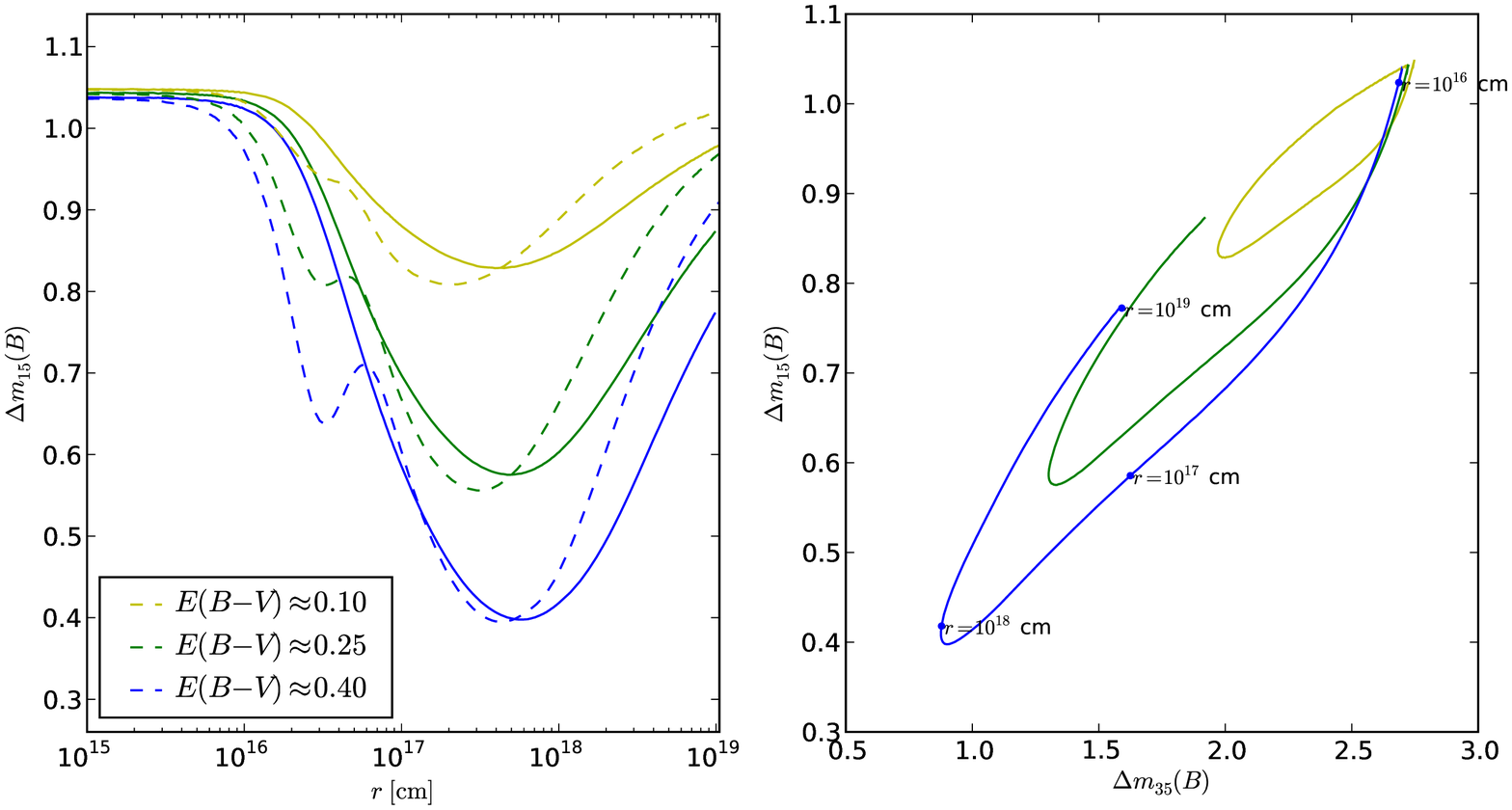}
  \caption{%
    {\em Left:} evolution of the lightcurve width expressed in terms of
    $\dbfifteen$ for cases (a) (dashed) and (b) (solid) for different 
    amounts of dust. {\em Right:} the fall time in terms of
    $\dbfifteen$ and $\dbthirtyfive$ as a function of the shell
    radius. %
    \label{fig:falltime}}
\end{figure*}

The most straight-forward method of quantifying post-maximum
lightcurve shape is in terms of the magnitude difference between
maximum and an arbitrary number of rest-frame days past maximum in a
given rest-frame passband. This method of measuring the fall-time was
introduced by \cite{1993ApJ...413L.105P} who chose the rest-frame
$B$~band and 15~days past maximum respectively, $\dbfifteen$. The
simulation results of the evolution of the $\dbfifteen$ parameter for
a few different CS-scenarios is presented in the far right panel of
Fig.~\ref{fig:rdependence} and in more detail in
Fig.~\ref{fig:falltime}.

For $10^{16}\lsim r \lsim 5\cdot 10^{17}$~cm, $\dbfifteen$ decreases
with $r$. However, for larger radii, the late photons populate the
tail of the lightcurve, \ie, and thus the original pristine lightcurve
fall-time at earlier epochs is gradually recuperated, while a plateau
is built up at late times. The significance and the variation of the
lightcurve tails in the examples shown suggests further studies of the
fall-time, also at epochs beyond day $15$. This is illustrated by the
right panel of Fig.~\ref{fig:falltime}, where two different radii for
the same dust configuration can give rise to the same value of
$\dbfifteen$, but different values at 35~days past maximum,
$\dbthirtyfive$. For very large radii, both the original values
$\dbfifteen$ and $\dbthirtyfive$ are recovered. For these extreme
radii, the imprint from the CS dust will be found for even later
epochs.

\begin{figure}
  \centering
  \includegraphics[width=.5\textwidth]{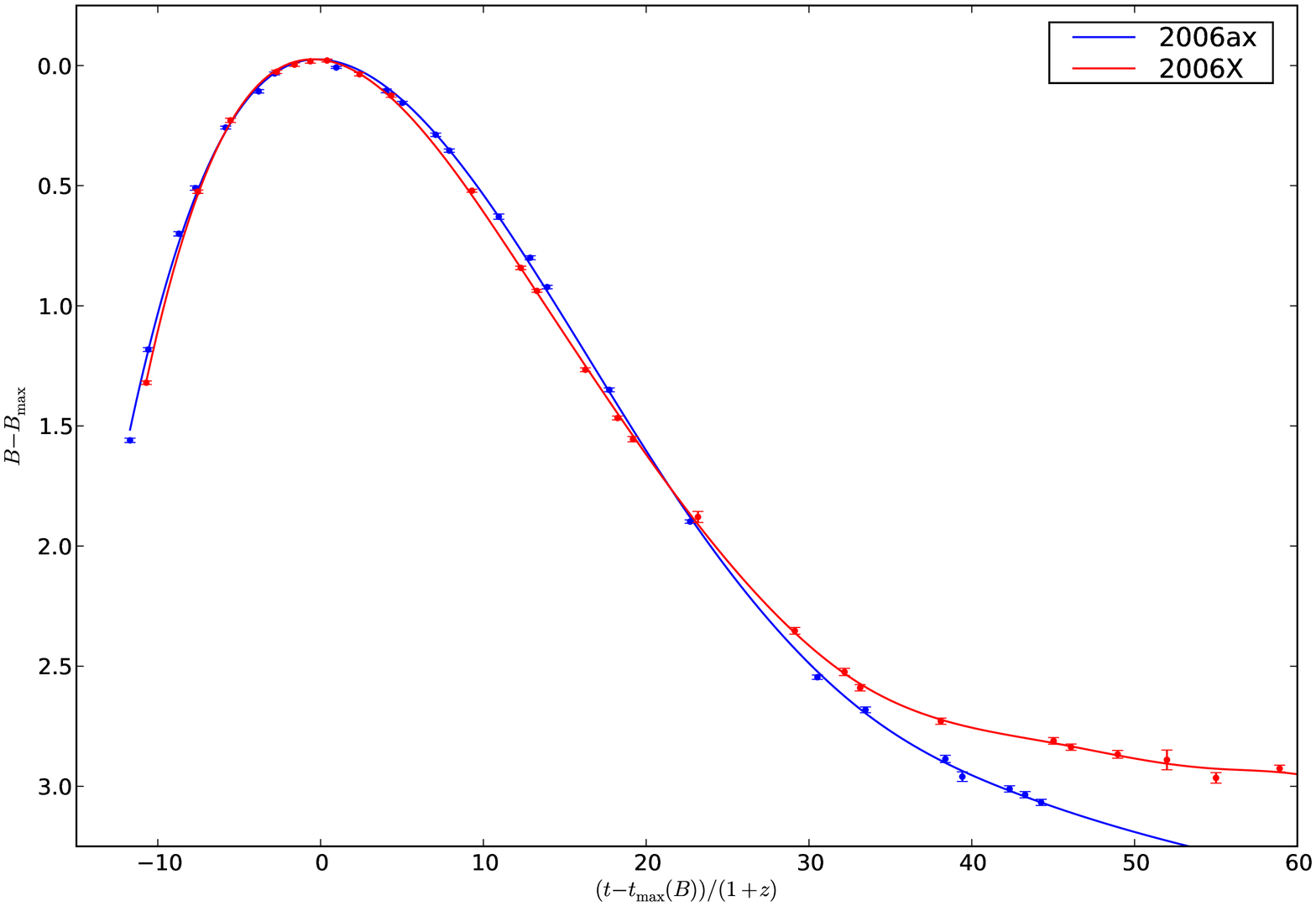}
  \caption{%
    $B$~band lightcurves of 2006ax and 2006X from the Carnegie
    Supernova Project with $\dbfifteen=0.99$ and $\dbfifteen=1.1$
    respectively. The solid lines are smoothed splines.  While the 
    fall-off between maximum and day~15 is steeper for 2006X
    than for 2006ax, this is not true between maximum and for e.g.
    day~30. The same qualitative appearance can be seen for the
    CS model in scenario~(b) in Fig.~\ref{fig:rdependence} if the
    shell radius is varied.
    \label{fig:Xax}}
\end{figure}

Another interesting effect occurs for the dust shell scenario (b) as
shown in Fig.~\ref{fig:rdependence}. The fall-off of the
$r=10^{18}$~cm curve is initially steeper than the $r=10^{17}$~cm
curve, but at day $\sim30$ past maximum, this changes and the
$r=10^{18}$~cm curve flattens out and cross the $r=10^{17}$~cm curve,
while the latter continues to fall. This qualitative behavior can
also be seen in real data: two \sneia normalized to the same maximum
brightness show different relative fall-off for different epochs, as
seen in Fig.~\ref{fig:Xax}.

\subsection{The rise-time -- fall-time relation}
\begin{figure}
  \centering
  \includegraphics[height=0.37\textwidth]{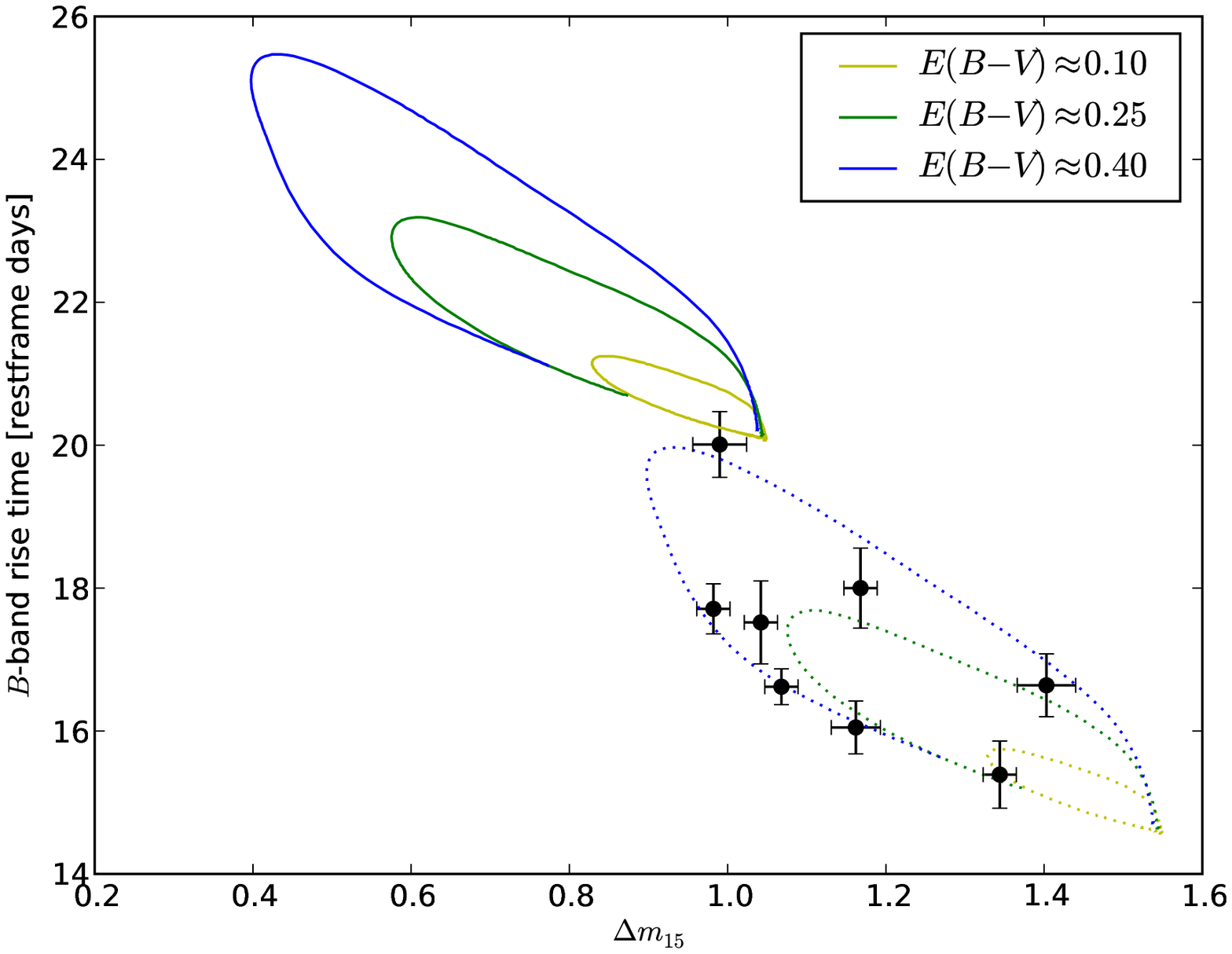}%
  \caption{Rise-time vs fall-time imposed by different CS-scenarios.
    The colors represents different amounts of CS-dust and
    different points on the individual curves represent varying
    dust radii. The plot is for a spherical scenario, \ie $r_i = 0$.
    Different values on $r_i$ give rise to similar size of
    the curves, but with slightly different shapes.
    \label{fig:strovink}}
\end{figure}

The impact of the time-delay effect from scattering on CS material is
expected to have a stronger impact on the fall-time than on the
rise-time of the lightcurve. The relation between the \sneia rise time
and $\dbfifteen$ has been studied both for high-quality data of
well-sampled nearby \sneia \citep{2007ApJ...671.1084S}, as well as in
a statistical approach for a much larger SDSS-II data set
\citep{2010ApJ...712..350H}. While \citet{2007ApJ...671.1084S} found
double-peaked distribution for the relation between rise and fall
time, this was not seen by \citep{2010ApJ...712..350H}.
Fig.~\ref{fig:rdependence} reveals that different CS-dust scenarios
could give rise to a relation between rise and fall time, even when
a single pristine light source is assumed. In
Fig.~\ref{fig:strovink} we show this relation for a few different
scenarios, together with the results from~\cite{2007ApJ...671.1084S}.

The absolute values of any lightcurve parameters determined from our
simulations will depend on the corresponding values of the pristine
source. Since the lightcurve parameters of the H07 template will
roughly reflect the mean values of the \snia population, our
simulations will only be able to produce lightcurves with longer rise
times and smaller $\dbfifteen$ than average. In the CS scenario it is
reasonable to assume that a ``naked'' \snia template have shorter
generic rise time and a faster fall-time, in order to compensate for
this we have also applied an add-hoc shift to all curves in
Fig.~\ref{fig:strovink} in order to match the data points. Further,
assuming that the generic $B$~band shape of a ``naked'' \snia is
similar to the H07 template that we use here, any modifications to the
relative lightcurve properties for different dust scenarios is likely
to be of second order, and the general trend and size of the effect
should still be valid.

The result from Fig.~\ref{fig:strovink} is that for \sne within a
narrow range of observed color excess, CS-dust could naturally give
rise to a bi-modal distribution in the relation between rise and fall
time for a small sample. However, if the accepted color range is
extended, the gap between the two populations will be filled. More
generally, it is intriguing that the rise-fall time scatter observed
in normal \sneia is compatible with our simulations.

\newcommand{\tbimax}{t_\mathrm{max}^B - t_\mathrm{max}^I}
\subsection{Time of maximum for different wavelengths}
\begin{figure}
\centering
\includegraphics[height=0.37\textwidth]{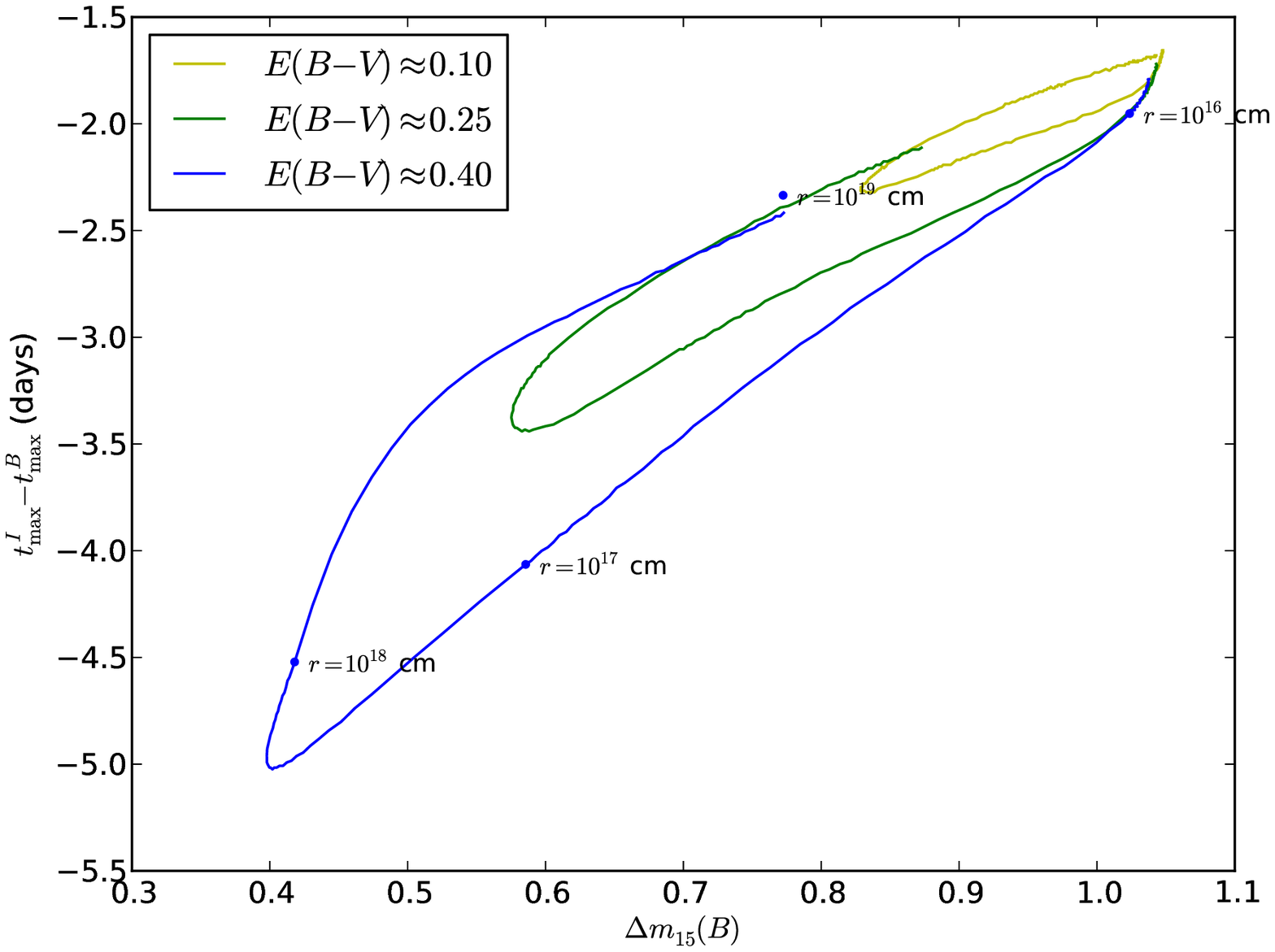}
\caption{The difference between the time of maximum in the $B$ and $I$
  bands for different amounts of CS LMC dust. The curves are shown for
  a dust sphere, case (a). For the one of the curves, a few different
  outer radii have been marked for guidance.}
\label{fig:tmaxshape}
\end{figure}

Up to this point we have focused on the effect of circumstellar dust
on the rest-frame $B$~band, the wavelength range that is traditionally
used for doing \sn cosmology. Since the scattering and absorption
cross-section decreases with wavelength, we expect the delay
distribution to affect bluer pass-bands more. One way of studying this
is to investigate the impact on the \sn rise-time for different
filters. However, this is related to comparing the time of maximum in
different bands, which is also a property that can be measured quite
accurately. Fig.~\ref{fig:tmaxshape} shows the effect on the
difference of the time of maximum between the $B$ and the $I$ bands,
$\tbimax$, for various CS scenarios. The general trend in the model
prediction is that the time between the $I$~band and $B$~band
lightcurve maxima increases for slow decliners. Furthermore, the
effect is quite sensitive to optical depth, i.e., to the color excess.


\section{Intrinsic color variations}
G08 explicitly describes both how circumstellar dust reddens the
pristine source and derives this reddening law. However, as we have
seen due to the time delay, mixing of photons originating from
different epochs is expected. Since the delay time, $\Delta t$,
depends on the shell radius, the observed color for a fixed dust
amount will have an intrinsic scatter with a magnitude depending on
the distribution of the shell size, $r$.

\begin{figure}
  \centering
  \includegraphics[width=.5\textwidth]{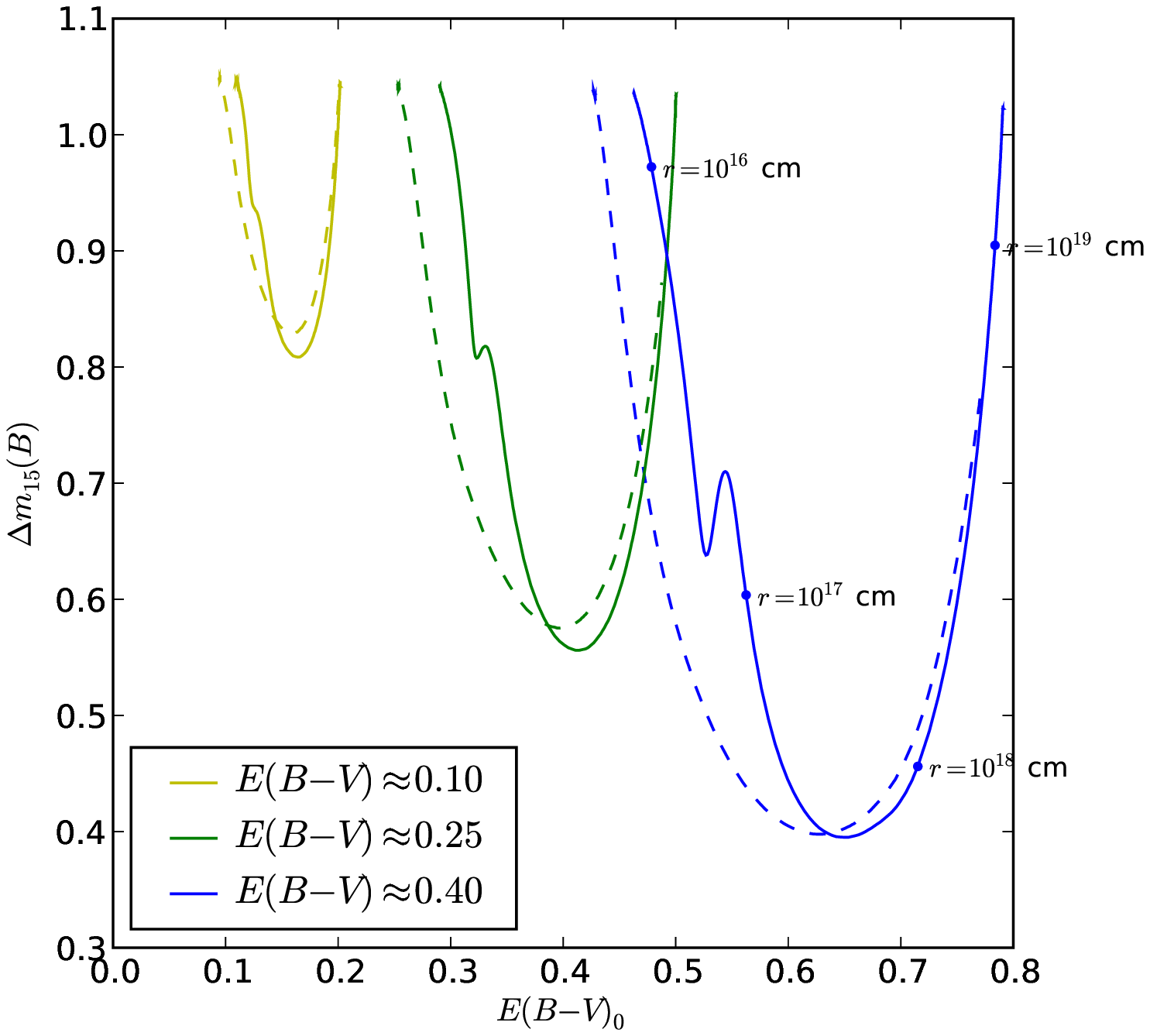}
  \caption{%
    The evolution of $\dbfifteen$ and the color excess at $B$-maximum
    as a function of the shell radius for cases (a) (dashed) and (b)
    (solid) respectively.
    \label{fig:shapevsc}}
\end{figure}

Fig.~\ref{fig:shapevsc} shows how a few different CS-scenarios could
affect both the lightcurve shape and the color at $B$~band maximum. As
with previous results we can see that the effects of both lightcurve
shape and color increases with the amount of CS-dust. There is also
degeneracy between different CS scenarios, so no trivial correlation
between fall-time and color excess is expected in the CS-scenario. The
CS-model thus naturally accommodates an intrinsic color scatter in
moderately reddened supernovae, $\sigma_{E(B-V)}\sim 0.05-0.1$ mag, in
agreement with observations
\citep{2008A&A...487...19N,2010ApJ...717...40K}.

\subsection{Time dependent color excess}
\begin{figure*}
\centering
\includegraphics[height=0.8\textwidth]{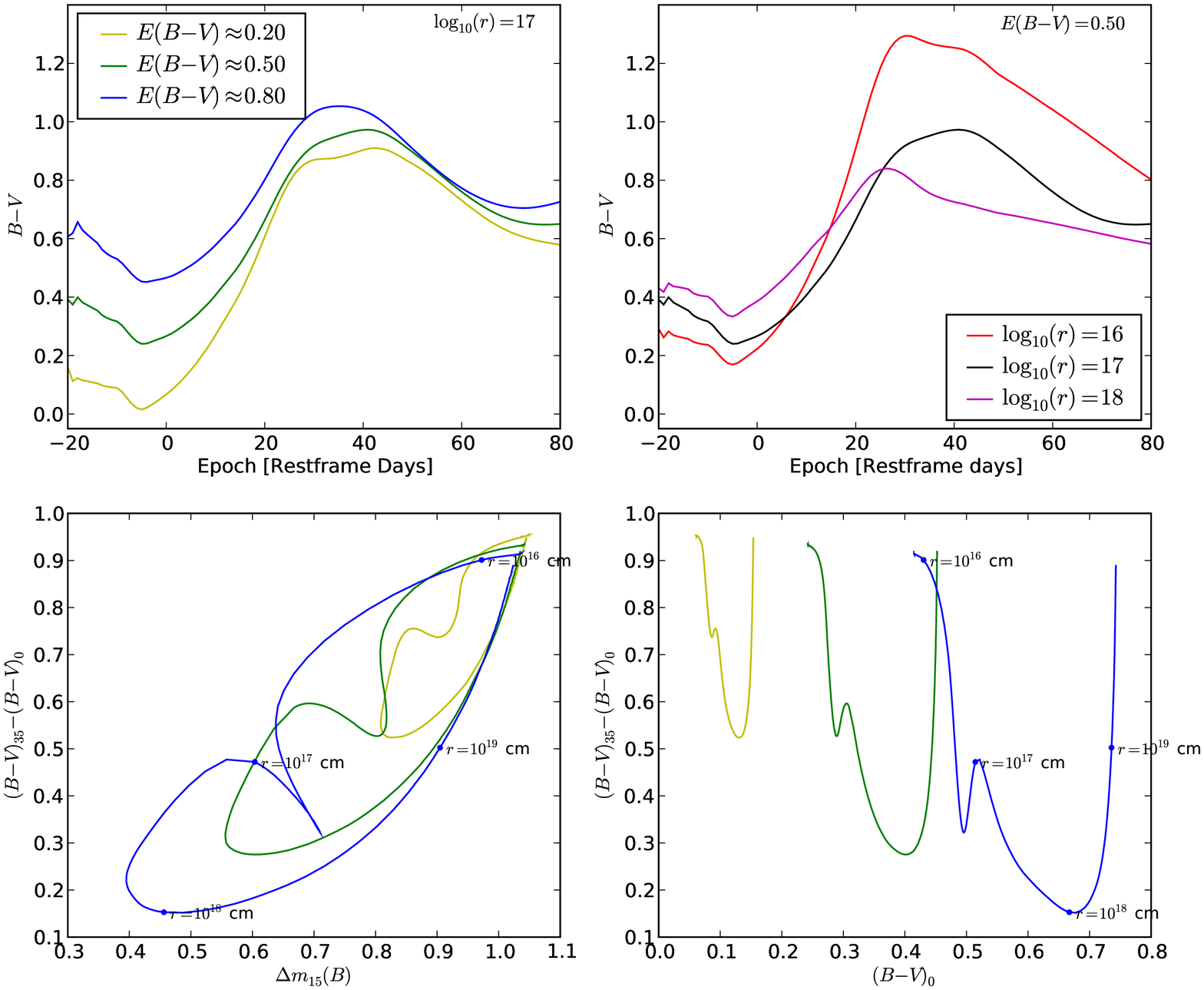}%
\caption{%
  The imposed color for different epochs.  The {\em upper} row shows
  the $B-V$ color evolution as a function of time, for different
  amounts of dust ({\em left}) and different radii ({\em right}).
  The lower {\em lower} row show the color difference between 35~days
  after $B$~band maximum and maximum brightness for different amounts
  of LMC dust.  This is shown versus the
  $B$~band fall-time ($\dbfifteen$) and the color at maximum in the
  {\em left} and {\em right} columns respectively.
  The simulated relation between the imposed lightcurve shape, 
  The dust radii for a one of the CS
  realizations have been plotted along the curve for guidance.  
  All results are shown for a CS scenario (a) with $r_i=0.95\cdot r$.  
  The legend describes the color excess the different CS scenarios
  would give rise to if the time-delay effect on the color is
  ignored.
  \label{fig:colorexcess}}
\end{figure*}

One of the salient features of the scenario in G08 is that blue
photons scatter more than red photons. This implies that blue photons
will, on average, be more delayed by multiple scattering since they
have a longer path-length before leaving the circumstellar dust
environment. This implies that the color excess due to the dust
component will acquire time dependence. With respect to a time
averaged color excess, the measured color will be redder pre-max
turning bluer after max, as shown in Fig.~\ref{fig:colorexcess}. This
effect should provide the most stringent bounds on the dust shell
sizes allowed by the data. \cite{2009PASJ...61..713Y} show in their
analysis of SN~2006X that once they correct for the color excess around
maximum light, the $V-R$ color is about 0.3 mag bluer than the normal
unreddened SN~2003du 2 months after lightcurve peak, in qualitative
agreement with the CS model prediction.

\section{Perturbations to the lightcurve shape-brightness relation}
Since lightcurve shape is the primary parameter to standardize type Ia
supernovae as distance indicators, it is important to investigate how
the relation between lightcurve shape and brightness may be affected
by the presence of circumstellar dust. As noted in Section
\ref{sec:sublimation}, an intrinsically brighter explosion will deplete
more of the surrounding dust, emphasizing the fainter-redder relation.
For shell radii $r\gsim 10^{16}$ cm, the optical lightcurve shape will
broaden the lightcurves potentially influencing the empirically
derived brighter-broader relation. Is it possible to obtain
a correlation between brightness and lightcure shape from the
interaction with the circumstellar environment? Although possible,
it invokes a somewhat contrived scenario: For shell sizes
$r\sim 10^{15}-5 \cdot 10^{17}$ cm, if the outwards velocity of the
circumstellar material, i.e., the white dwarf wind velocity, correlates
with the amount of $^{56}$Ni powering the explosion, brighter \sne would
have large $r$ and become slow decliners. We are not aware of any
supernova models that have predictions in this respect.

More generally, if the lightcurve shape is a combination of effects,
involving both physics of the progenitor system and the material
around the white dwarf, then one would expect a residual dependence of
brightness on lightcurve shape, even after the main component has been
calibrated out. In particular, we note that the optical lightcurve
shape beyond 30 days after $B$~band peak, will be noticeably affected
for a wide range of shell sizes.

\section{The secondary ``bump'' in the $I$~band lightcurve}
The type Ia $I$~band lightcurve  
exhibits a secondary maximum approximately 15-30 days after the $B$~band 
lightcurve maximum which, in turn, typically happens within 2 days  
from the primary $I$~band peak \citep{2005A&A...437..789N}. The time
gap between the two $I$~band peaks, as well as the relative strength
of the secondary peak has been shown to correlate with the $B$~band
lightcurve shape \citep{2005A&A...437..789N}.

\begin{figure*}
\centering
\includegraphics[width=\textwidth]{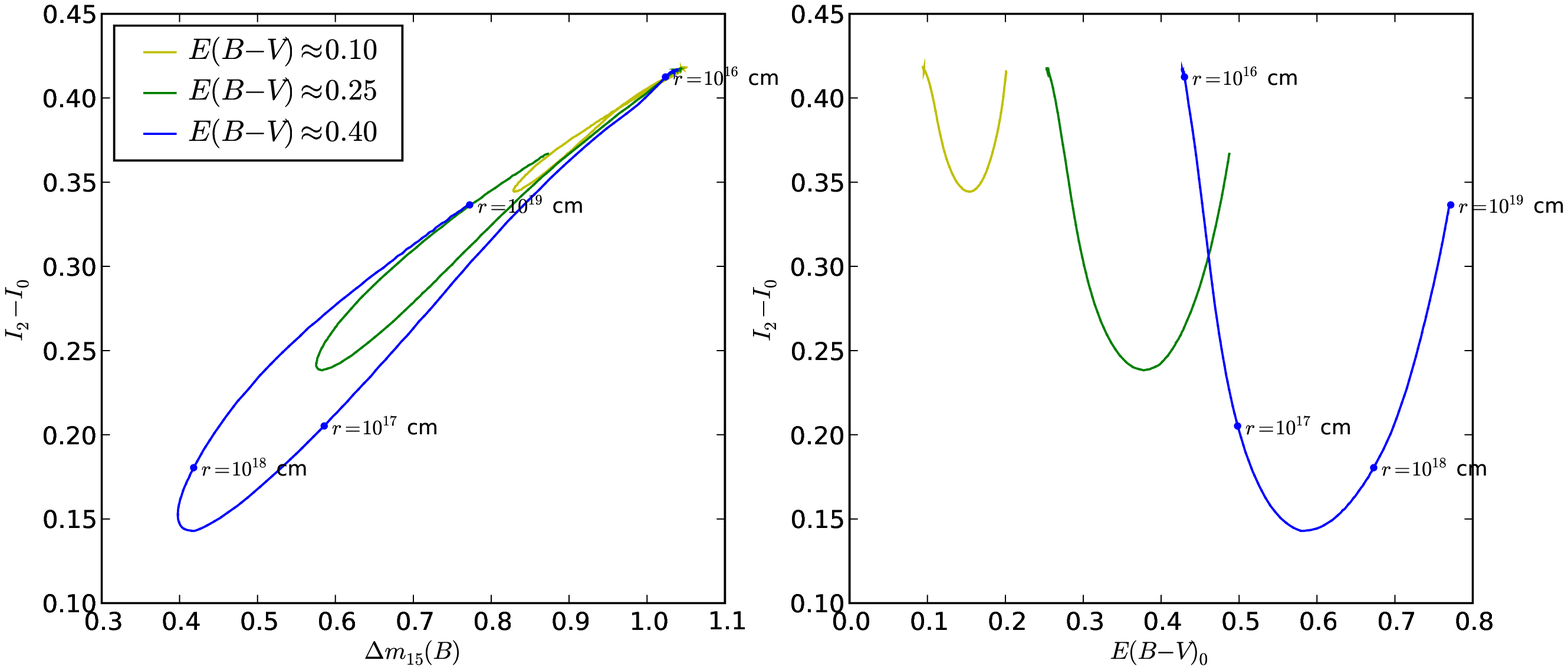}
\caption{%
  The imposed relation between the extreme points of the $I$~band
  lightcurve. The difference in brightness between the second and
  first peak is shown vs the fall-time ($\dbfifteen$) and the color
  excess at $B$-max ($E(B-V)_0$) respectively. The results are shown
  for a CS model with $r_i=0.00\cdot r$ and LMC-type dust. %
  \label{fig:iband}}
\end{figure*}

The implication of the imposed time-delay from scattering on
circumstellar dust on the $I$~band peaks is shown in
Fig.~\ref{fig:iband}. We note that for a constant shell size we expect
the brightness difference (upper row) between the peaks, as well as
the significance of the secondary peak, to decrease with
an increased amount of dust. The effect comes from the fact that
photons from the first peak are shifted to both fill the gap between
two peaks and emphasize the second bump, and is therefore also
correlated with $B$~band fall-time and $E(B-V)$.

If the shell size is allowed to vary, the usual degeneracy between
shell size and the amount of dust is obtained for the observables.
Despite this degeneracy, it is interesting to note that for a broad
range of dust properties and shell sizes, a relation between the
$I$~band lightcurve properties and $E(B-V)$ is expected.

\section{Spectral features}
\cite{2010AJ....139..120F} found that different fitted reddening
corrections were obtained for their data set depending on whether the
reddest \sne were included in the fit or not. For the full data set
they obtained $R_V=1.7$, while they fitted $R_V=3.2$ if the
high-reddened \sne were excluded. In contrast to this,
\cite{2010ApJ...716..712A} found that redder \sne in the Union2 sample
prefer a higher color correction than bluer \sne and speculate that
the color correction may very well be more complex than a simple
linear relation.

Further, \cite{2009ApJ...699L.139W} found a bi-modality in the
color-magnitude relation over a broad color range, with the two groups
preferring different values of $R_V$ independently of the color. They
also observed a correlation between the velocity of
Si~II($\lambda6355$) and the fitted value of $R_V$, concluding that
objects with high velocity tend to prefer a lower value of $R_V$ than
\sne with low velocity.

One possible explanation for the bi-modality could be that the color
of the low-$R_V$ \sne primarily originates from CS dust reddening,
while the high-$R_V$ objects are dominated by extinction from
interstellar dust. 

\begin{figure*}
  \centering
  \includegraphics[width=0.8\textwidth]{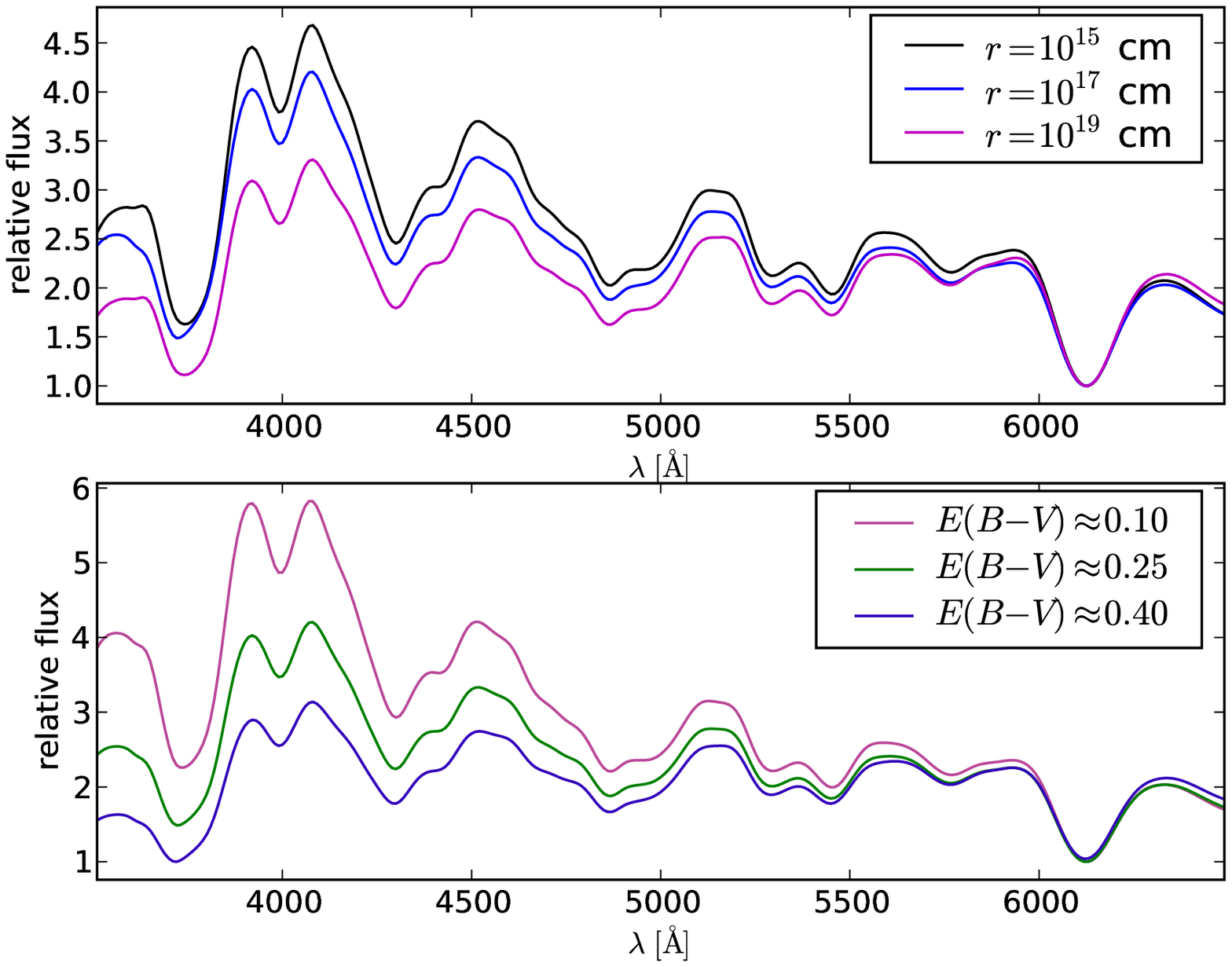}
  \caption{%
    The the impact on the spectrum at maximum from CS dust.  The {\em
      upper} panel shows the evolution for different shell radii,
    assuming LMC dust that would give $E(B-V)\approx0.25$ in the limit
    when the radius goes to zero.  The {\em lower} panel shows the
    spectrum for different amounts of dust when the radius is fixed to
    $r=10^{17}$~cm.  The inner radius is fixed to $r_i = 0.95\cdot r$
    for all cases. %
  \label{fig:silicon}}
\end{figure*}

The time-delay of photons from CS dust scattering is expected to blend
the spectral energy distributions between different epochs, \ie a
spectral feature for a given epoch will effectively be a superposition
of the feature for all previous epochs. Since time-delay distribution
varies with wavelength, the amount of blending will also decrease with
wavelength. Furthermore, since the minima of the supernova absorption
features trace the velocity of the receding photosphere during the
first weeks after the explosion, we expect that delayed photons
arriving around maximum light will originate from a region further out
in the photosphere, i.e., at higher velocities. However, our
simulations indicate that these effects are small (see
Fig.~\ref{fig:silicon}), typically accounting for velocity variations
of a few hundred km~s$^{-1}$, i.e., compatible with the velocity
scatter between measurements of normal \sneia, but not enough to
explain significant outliers like SN~2006X \citep{2009PASJ...61..713Y}.

%

\section{Summary and conclusions}
Dust layers surrounding the progenitor systems of \sneia
could help explain the observed reddening law. In this work, we
have performed Monte Carlo simulations that indicate that the scenario
with circumstellar dust would also perturb the supernova optical lightcurves
in a manner that resembles empirical findings, e.g., an ``intrinsic''
color variation of $\sigma_{E(B-V)} \sim 0.05-0.1$ arises naturally
in our simulations.  Since multiple
scattering can only introduce a delay in time, the net effect is a
broadening of the lightcurve. While the broadening increases
monotonically with the density of dust, the relation with the dust
shell size is less straight forward: there is an increase for outer radii
about $\sim10^{17}$ cm, while for even larger radii, the late photons
populate the tail instead, and the lightcurve width is closer to the
original one. We conclude that well sampled multi-band lightcurves
of near-by \sneia reaching 2-3 months after lightcurve peak would be
critical to probe the existence of dust in the circumstellar environment.
Our simulations also suggest that for large optical depths,
the lightcurve shape perturbation should be notably different before
and after maximum. Furthermore, unlike the case for reddening by interstellar
dust, color excess should be epoch dependent: redder earlier on and bluer
after maximum. 


\edit{%
  A strong case for the CS scenario would be if the observed lightcurve
  perturbations in the optical could be combined with a detection of the
  emission at longer wavelengths from the heated CS material. A model
  for this emission was constructed by \cite{1983ApJ...274..175D} for
  circumstellar dust around Type~II SNe. Although the exact properties
  of the emission is highly model dependent, it is generally expected to
  last after the optical light has faded.}





\acknowledgments
We are grateful to the organizers and participants of the Aspen 2010
\snia workshop for helpful discussions at the time parts of this manuscript
was written. We also acknowledge Adolf Witt for helpful comments on the 
mansucript.

\renewcommand{\theequation}{A\arabic{equation}}
\renewcommand{\thefigure}{A\arabic{figure}}
\setcounter{equation}{0}  
\setcounter{figure}{0}  

\appendix
\renewcommand\thesection{Appendix \Alph{section}}

\section{A spherical dust shell}\label{sec:sphere}
\begin{figure}[ht]
  \centering
  \includegraphics[width=\textwidth]{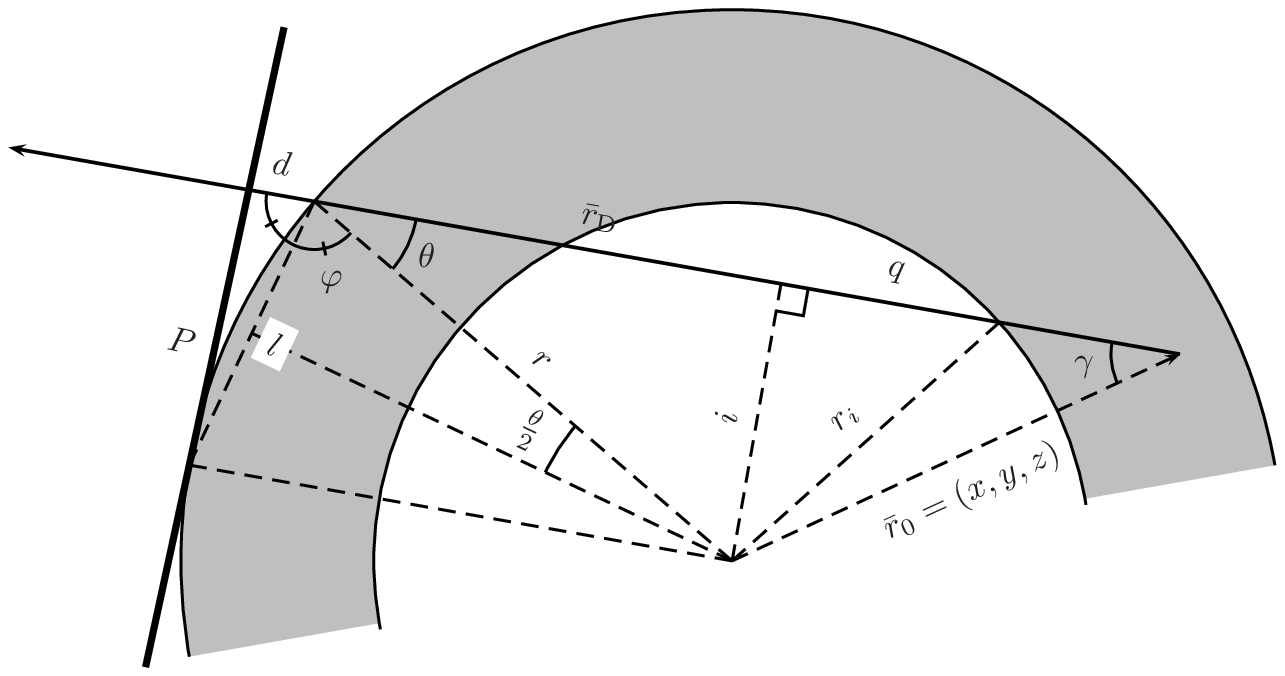}
  \caption{Illustration of a scattered photon (\emph{thin solid line})
    propagating through dust shell. The figure shows the intersecting
    plane of the sphere that is defined by the two vectors $\posvec$
    and $\disvec$.\label{fig:sphere}}
\end{figure}

A simple model of a spherical dust-shell with radius, $r$, and inner
radius, $r_i = R\cdot r$, where $0<R<1$ is illustrated in
Figure~\ref{fig:sphere}. Also shown in the figure is a photon leaving
the sphere (solid line).
The current position of the photon is defined by the vector, $\posvec
= (x,y,z)$, and the next position, $\nexvec$, is given by $\nexvec =
\posvec+\disvec$, where $\disvec$ is the displacement vector $\disvec
= (dx,dy,dz)$.

\paragraph{Photon traveling inside the shell} The impact parameter,
$i$, is the minimum distance to the center of the sphere for any given
photon-path, and can be defined as.
\[
i = \left\{%
  \begin{array}{ll}
    \poslen\cdot\sin\gamma & \textrm{if } 0 \leq \gamma \leq \frac{\pi}{2}\,,\\
    \poslen                & \textrm{if } \frac{\pi}{2} < \gamma\,.
  \end{array} \right.
\]
Here, $\gamma$ is given by
\begin{eqnarray*}
  \posvec\cdot\disvec & = & \poslen\dislen\cos(\pi-\gamma) = 
  x\cdot dx + y\cdot dy + z\cdot dz\\
  \cos\gamma & = & -\cos(\pi-\gamma) = -\frac{x\cdot dx + y\cdot dy + z\cdot dz}{%
    \poslen\dislen}\, .
\end{eqnarray*}
If $i < r_i$, the photon will cross the inner radius of the shell and
then re-enter the shell. This will in turn extend the mean free path
of the photon by the amount $2q = 2\sqrt{r_i^2 + i^2}$, which is the
length of the path within $r_i$.

\paragraph{Path length} For each distance between interactions,
$\dislen$ is added to the total distance traveled by the photon
before leaving the sphere. For a scattered photon leaving the sphere,
the last distance, $s$, traveled inside the sphere is given by
\[
  s^2 = r^2 + \poslen^2 - 2r\poslen\cos\left(\pi-\theta-\gamma\right)\,,
\]
where $\cos\theta = \sin\gamma \cdot \poslen/r$. The distance
traveled by a scattered photon should be compared to the distance
traveled by a non-interacting photon, which means that the distance,
$d$, between the surface of the sphere and the plane (thick, solid
line in Figure~\ref{fig:sphere}) perpendicular to $\disvec$ at
distance $r$ from the center must be added to the total distance,
where $d$ is given by
\begin{eqnarray*}
  l & = & 2r\cdot\sin\frac{\theta}{2}\\
  \varphi & = & \pi - \frac{\pi}{2} - \frac{\theta}{2} =
  \frac{\pi-\theta}{2}\\
  d & = & l\cdot\cos\varphi =
  2r\cdot\sin\frac{\theta}{2}\cos\frac{\pi-\theta}{2} =
  2r\cdot\sin^2\frac{\theta}{2}\, .
\end{eqnarray*}
The total distance traveled, $D$, can then be summarized as
\begin{equation}
  D = \left\{%
      \begin{array}{ll}
    r & \textrm{for non-interacting photons}\,,\\
    \sum {\dislen}_n + s + d & \textrm{for scattered photons}\,.
  \end{array}\right.
\label{eq:pathlength}
\end{equation}

\bibliography{ms}

\clearpage




\clearpage

\clearpage



\end{document}